# PRESSURE-INDUCED INSULATING STATE IN $(La, Sr)CoO_3$

R. Lengsdorf[1], M. Ait-Tahar[1], S. S. Saxena[2], M. Ellerby[3], D. I. Khomskii[1], H. Micklitz[1], T. Lorenz[1], and M. M. Abd-Elmeguid[1]

[1] II. Physikalisches Institut, Universität zu Köln, Zülpicher Str. 77, 50937 Köln, Germany

[2] Cavendish Laboratory, University of Cambridge, Madingley Road, Cambridge CD30HE, UK

[3] Department of Physics and Astronomy, University College of London, Gower Street, London WC IE6BT, UK

We have investigated the effect of pressure on the electronic, magnetic and structural properties on single crystal of conducting, ferromagnet ($T_C$ = 157 K) $La_{0.82}Sr_{0.18}CoO_3$ located near the boundary of the metal-insulator transition. Contrary to the results reported on related systems, we find a transition from the conducting state to an insulating state and a decrease of $T_C$ with increasing pressure while the lattice structure remains unchanged. We show that this unusual behavior is driven by a gradual change of the spin state of $Co^{3+}$ ions from magnetic intermediate-spin $(t_{2g}^5 e_g^1; S = 1)$ to a nonmagnetic low-spin $(t_{2g}^6 e_g^0; S = 0)$ state.

The study of the electronic and magnetic properties of strongly correlated transition metal oxides $(TMO)$, such as perovskites of the type $RMO_3$ ($R$ = rare earth ion, $M$ = transition metal) near the metal-insulator (MI) transition has recently attracted considerable attention. This is due to the fact, that in this class of systems the MI transition is driven by strong correlation effects associated with electron-electron interaction and the interplay between charge, orbital and spin degrees of freedom. The MI transition can be induced by varying the carrier concentration, temperature, magnetic field and internal or external pressure [1]. Thus, studying such an interplay is a fundamental issue for a better understanding of the nature of the MI transition as well as the novel phenomena observed in these systems e.g. colossal magnetoresistance or even high temperature superconductivity.

In this respect, external pressure can be very effective, e.g. by modifying the effective bandwidth ($W$) of the transition metal by changing the $M-O$ bond length ($d_{M-O}$) and/or the $M-O-M$ bond angle ($\theta$) thereby providing a unique tool to tune electronic and magnetic properties by "bandwidth control" of these systems. In manganese perovskites $R_{1-x}A_xMnO_3$ ($A = Ca, Sr, Ba$), for example, it has been shown that increasing pressure decreases $d_{Mn-O}$ and increases $\theta$. Both variations lead to an increase of $W$ and thereby stabilize the ferromagnetic metallic state [2-4]. Regarding the charge degree of freedom, recent high pressure studies on $Sr_{2/3}La_{1/3}FeO_3$ show a pressure-induced transition from the charge-disproportionate antiferromagnetic insulating state to a charge-uniform ferromagnetic metallic state [5].

Transition metal oxides containing $Co^{3+}$ ion are of special interest, because in addition to the usual charge, spin and orbital degrees of freedom they possess an extra degree of freedom, namely the possibility to change the spin-state of the $Co^{3+}$ ion. This can occur in $Co^{3+}$ $TMO$ compounds if the crystal field energy $\Delta_{CF}$ and the intraatomic exchange energy $E_{ex}$ (Hund's rule coupling) are comparable. Such a delicate energy balance leads to spin-state transitions which can be induced by changing the temperature or composition [e.g. 6–8]. Among these systems, the rhombohedral perovskites $La_{1-x}Sr_xCoO_3$ ($0 \leq x \leq 0.5$) represent a unique system which allows one to investigate the interplay between the spin-state degree of freedom and electronic and magnetic properties particularly near the MI transition. In $LaCoO_3$ the ground state of $Co^{3+}$ ions is, contrary to typical Mott insulators, nonmagnetic with a low-spin (LS) configuration $(t_{2g}^6 e_g^0; S=0)$, and these ions can be thermally excited to an intermediate-spin (IS) state $(t_{2g}^5 e_g^1; S=1)$ at $T \approx 100$ K [9-11]. At higher temperatures around 500 K the system undergoes a MI transition [12]. However, by doping with $Sr^{2+}$, the ground state becomes ferromagnetic for $x \geq 0.18$ (through a spin-glass like region between $0 < x < 0.18$); at the same time the electrical conductivity increases with increasing $x$ and for $x \approx 0.2$ the system undergoes a transition to a metallic state [13-16,17]. It is generally accepted that replacing $La^{3+}$ by $Sr^{2+}$ creates formally $Co^{4+}$ ions and that the double exchange between $Co^{4+}$ and the remaining $Co^{3+}$ leads to a ferromagnetic coupling [15,18]. However, little is known about the correlation between the spin- and charge degree of freedom near the MI transition, i.e. whether and to what extent these spin-state transitions affect the MI transition.

In the present work, we have investigated the effect of pressure on the electronic, magnetic and structural properties of a conducting, ferromagnetic single crystal of $La_{0.82}Sr_{0.18}CoO_3$ located near the boundary of the MI transition. We found, contrary to the results reported on related systems ($La_{1-x}Sr_xMnO_3$, $La_{1-x}Sr_xFeO_3$ and $RNiO_3$), a dramatic pressure-induced *increase* of the electrical resistivity (~ 4 orders of magnitude) and a decrease of $T_C$ with increasing pressure, while the lattice structure remains unchanged. This is quite unusual because in all other correlated oxides pressure always leads to an increase of $3d$ bandwidth and thereby to a transition to a more conducting state. The opposite effect observed in this work is rather unique and is apparently connected with different physical mechanisms. We explain this behavior as a consequence of a change of the $Co^{3+}$ spin state from a magnetic IS configuration to a nonmagnetic LS configuration under pressure.

Single crystals of $La_{0.82}Sr_{0.18}CoO_3$ were grown by the travelling zone method in a 4-mirror image furnace under oxygen pressure of 5 bar. Details of the preparation and characterization are published elsewhere [17]. The pressure dependence of the electrical resistivity up to 14 GPa and for 4.2 K $< T <$ 300 K was measured in a Bridgman-type high pressure cell. This setup is basically similar to that described in Ref. [19]. Sintered diamond anvils with a flat surface of 4 mm diameter were used to achieve pressure up to 14 GPa. A sample of 25 x 140 x 280 (µm³) was measured by conventional four-terminal method. Steatite was used as a pressure transmitting medium and the pressure gradient within the pressured cavity was between 5 – 7 %. The pressure dependence of the magnetization up to 1.5 GPa was measured in an external magnetic field of 2 T and between 2 K $< T <$ 300 K by means of a SQUID magnetometer. A miniature pressure clamp (teflon cell) was used as described elsewhere [20]. The pressure dependence of the lattice parameters up to 15 GPa was measured on powdered samples by energy dispersive x-ray diffraction at HASYLAB using the diamond anvil cell technique.

Fig. 1 shows the temperature dependence of the resistivity $\rho(p,T)$ at different pressures up to 5.7 GPa. The $\rho(T)$-curve at ambient pressure shows a local maximum at a temperature $T_{max} \approx$ 138 K (s. inset of Fig. 1) slightly below the transition temperature $T_C$ to a ferromagnetic state ($T_C \approx$ 157 K at ambient pressure) as deduced from magnetization measurements [17]. As is evident from Fig. 1, with increasing pressure up to 5.7 GPa the electrical resistivity *increases dramatically* by about 4 orders magnitude, indicating a strong reduction of electron hopping with increasing pressure. By increasing the pressure above 5.7 GPa (s. Fig. 2) we find no further increase of $\rho(T)$; instead at $T$ = 4.2 K it decreases by about 30 % at 14 GPa compared with the corresponding value at 5.7 GPa. The pressure dependence of $\rho(p)$ at 4.2 K is shown in the inset of Fig. 2. The maximum of $\rho(T)$, $T_{max}$, which is correlated with the value of $T_C$, is shifted to lower temperatures with increasing pressure (s. Fig. 2). We obtain a decrease of $T_{max}$ of about 8 K between $p$ = 0 and 1.9 GPa [21]. Assuming that this decrease of $T_{max}$ is equal to that of $T_C$, we estimate a pressure-induced relative decrease of $T_C$ which amounts to $\frac{\partial \ln T_C}{\partial p} \approx -2.6 \cdot 10^{-2} GPa^{-1}$.

The observation of a strong increase of the electrical resistivity (i.e. suppression of electron hopping), and reduction of $T_C$ with increasing pressure is unexpected and opposite to the results reported on related $TMO$ $La_{1-x}Sr_xMnO_3$ [3] (and $La_{1-x}Sr_xFeO_3$ [5]). For example, in $La_{0.75}Sr_{0.25}MnO_3$ one finds an enhancement of electron hopping with increasing pressure that stabilizes the ferromagnetic metallic state via "double exchange". The relative increase of $T_C$ with pressure in this system amounts to $\frac{\partial \ln T_C}{\partial p} \approx +6.4 \cdot 10^{-2} GPa^{-1}$.

In the following we discuss two possible explanations for the observed suppression of electron hopping and reduction of $T_C$ in $La_{0.82}Sr_{0.18}CoO_3$ with increasing pressure for p ≤ 5.7 GPa. The first possibility could be a pressure-induced structural phase transition and/or an unusual change of the $Co-O-Co$ bond angle and $Co-O$ bond length. In Fig. 3 we show the pressure dependence of the lattice parameters $a$ and $c$ of the hexagonal unit cell as obtained from our diffraction measurements at 300 K. As shown in Fig. 3, we find within the accuracy of measurements no discontinuity in the pressure dependence of the lattice parameters $a$ and $c$ (and the volume), thus showing no indication of a structural phase transition. The value of the bulk modulus $(B_0)$ and its derivative $(B_0^{'})$ as obtained from the fit of the equation of state using the Birch-Murnaghan equation, $B_o = 158(8) GPa$ and $B_0^{'} = 5(1)$, are found to be close to the values for $LaCoO_3$, $B_o = 150(2) GPa$ and $B_0^{'} = 4$ [22].

From our x-ray diffraction data it is difficult to deduce information about the pressure dependence of the $Co-O$ bond length and $Co-O-Co$ bond angle. However, such information can be obtained from recent high resolution angle dispersive x-ray diffraction data on $LaCoO_3$ [22]. Here, it is shown that with increasing pressure the $Co-O$ bond length decreases more rapidly than that of La-O, resulting in a large increase of the $Co-O-Co$ bond angle (from 166° to ≈ 177° at ≈ 4 GPa). The increase of the bond angle with increasing pressure is the usual structural response of $TMO$ with perovskite structure and is known to enhance electron hopping, thereby stabilizing the metallic state. We observe exactly the opposite response to pressure, finding instead a pressure-induced suppression of electron hopping and reduction of $T_C$ with elevated pressure. Therefore, we have to discuss another explanation, not based on structural changes: the possibility of a pressure-induced spin-state transition from the magnetic IS $Co^{3+}$ state to a nonmagnetic LS state. This is based on two experimental facts: first, the ionic radius of LS $Co^{3+}$ (0.545 Å) is smaller than that of HS $Co^{3+}$ (0.61 Å) [23,24]. The second being the energy of the crystal field splitting $\Delta_{CF}$ in $LaCoO_3$ which has been found to increase remarkably with increasing pressure [25]. Thus, pressure is expected to favor the population of the LS state at the expense of depopulation of the IS state. A direct consequence of such a pressure-induced transition from the magnetic IS to a nonmagnetic LS state in $La_{0.82}Sr_{0.18}CoO_3$ would be a reduction of the magnitude of the $Co$ saturation moment with increasing pressure. As the LS $LaCoO_3$ compound is known to be an insulator, one can expect that this pressure-induced IS to LS transition in $La_{0.82}Sr_{0.18}CoO_3$ would also suppress the conductivity. This is what we observed experimentally.

In order to verify this interpretation we have measured the pressure dependence of the saturation magnetization of $Co$ $(\mu_{Co})$ in $La_{0.82}Sr_{0.18}CoO_3$. Fig. 4 displays the temperature dependence of the magnetization in an external magnetic field of 2 T for ambient pressure and 1.0 GPa. The decrease of the magnetization at low temperatures with increasing pressure is evident from this figure. We obtain from the experimental data a decrease of $\mu_{Co}$ (at 5 K) from 1.11(1) $\mu_B$ at ambient pressure to 1.05(1) $\mu_B$, i.e. 5.4(2) % at 1 GPa. This corresponds to a decrease of $\mu_{Co}$ by more than 30 % at 5.7 GPa, if one assumes a linear decrease of $\mu_{Co}$ with increasing pressure. In addition, we find a decrease of $T_C$ ($\approx$ 4 K/GPa) which is similar to that estimated from our high pressure resistivity data (s. above). It is clear, that the observed decrease of $T_C$ with increasing pressure is a consequence of the reduction of the $Co$ magnetic moment. These experimental findings strongly support our suggestion of a pressure-induced IS to LS transition in $La_{0.82}Sr_{0.18}CoO_3$.

In the following we present a qualitative model explaining why such a spin-state transition results in a suppression of electron hopping. Assuming that at ambient pressure the ground state of $La_{0.82}Sr_{0.18}CoO_3$ is a homogenous mixture of IS $Co^{3+}$ and LS $Co^{4+}$ ions [15,17,18], we see that the conduction occurs predominantly through hopping of an $e_g$ electron from $Co^{3+}$ $(t_{2g}^5 e_g^1)$ to $Co^{4+}$ $(t_{2g}^5 e_g^0)$. Furthermore, this transport process can account for the magnetic coupling of the localized $t_{2g}$ electrons (double exchange), resulting in a ferromagnetic conducting state. This is demonstrated in Fig. 5. With increasing pressure the energy of the crystal field splitting increases $(\Delta'_{CF} > \Delta_{CF})$ which results in a gradual depopulation of the IS $Co^{3+}$ state and its crossover to a LS state. At sufficiently high pressures ($p \approx 5.7$ GPa) the IS state will be largely depopulated and a large part of the ground state of $Co^{3+}$ is then predominantly a LS state $t_{2g}^6 e_g^0$ configuration. Consequently, the relatively strong $e_g$ hopping at ambient pressure between $Co^{3+}$ and $Co^{4+}$ would be strongly suppressed and only a very weak $t_{2g}$ hopping would remain. Thus, the gradual depopulation of the IS $Co^{3+}$ state with increasing pressure can explain the observed suppression of electron hopping in $La_{0.82}Sr_{0.18}CoO_3$.

In this context, we would like to emphasize that the effect of the pressure-induced IS to LS transition on the electron hopping *prevails over* the opposite effect of an increase of the $Co-O-Co$ bond angle with increasing pressure. This clearly shows that the strong correlation between the spin-state and charge degrees of freedom in $(La,Sr)CoO_3$ dominates the MI transition.

Finally, we discuss the observed decrease in the electrical resistivity with increasing pressure above 5.7 GPa (Fig. 2). As mentioned before, the resistivity decreases by about 30 % at 4.2 K between 5.7 GPa and 14 GPa. This decrease can be explained in the following way: above $\approx 5.7$ GPa a large part of the $Co^{3+}$ is already transformed from the IS to the LS state, so that this mechanism is not efficient any more and the usual tendency due to broadening of the effective bandwidth with increasing pressure prevails. This leads to a consequent decrease of the electrical resistivity.

In conclusion, we have investigated the effect of pressure on electronic, magnetic and structural properties of the ferromagnetic conducting perovskite $La_{0.82}Sr_{0.18}CoO_3$. We observed a quite unusual effect: a transition from the *conducting state to an insulating state*, which is contrary to expectations. We explain this unusual behavior as a consequence of a pressure-induced transition from the magnetic intermediate-spin state to a nonmagnetic low-spin state of $Co^{3+}$. This interpretation is supported by the observed reduction of the $Co$ saturation magnetic moment with increasing pressure and provides a natural explanation of the observed effect. Our results demonstrate the very important role of spin-state degree of freedom for the MI transition in materials with correlated electrons.

M. M. A., D. I. K. and T. L. thank L. H. Tjeng for fruitful discussions. This work was supported by the Deutsche Forschungsgemeinschaft through SFB 608.

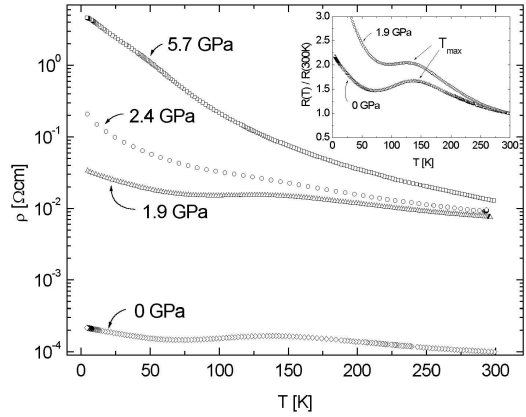

Fig. 1: Temperature dependence of the resistivity $\rho(p,T)$ of $La_{0.82}Sr_{0.18}CoO_3$ at different pressures up to 5.7 GPa. Inset shows the anomaly around $T_C$ at ambient pressure and 1.9 GPa.

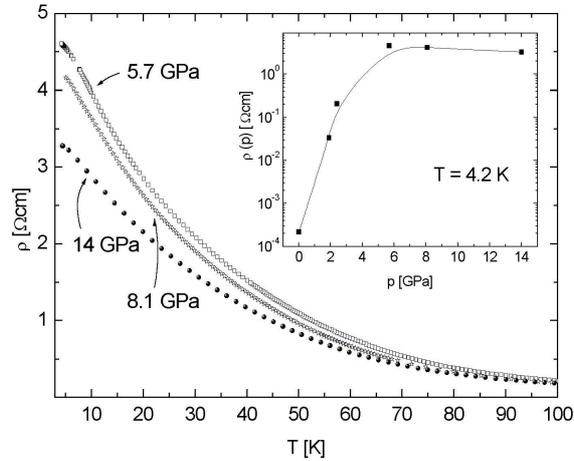

Fig. 2: Temperature dependence of the electrical resistivity $\rho(p,T)$ of $La_{0.82}Sr_{0.18}CoO_3$ in the pressure range 5.7 GPa $\leq p \leq$ 14 GPa. Inset shows the values of $\rho(p)$ at $T = 4.2$ K as a function of pressure in the whole pressure range.

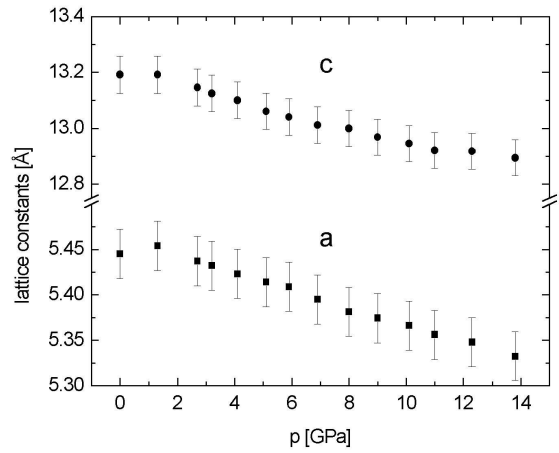

Fig. 3: Pressure variation of the lattice parameters $a$ and $c$ of the hexagonal unit cell of $La_{0.82}Sr_{0.18}CoO_3$ as obtained from energy dispersive x-ray diffraction at 300 K.

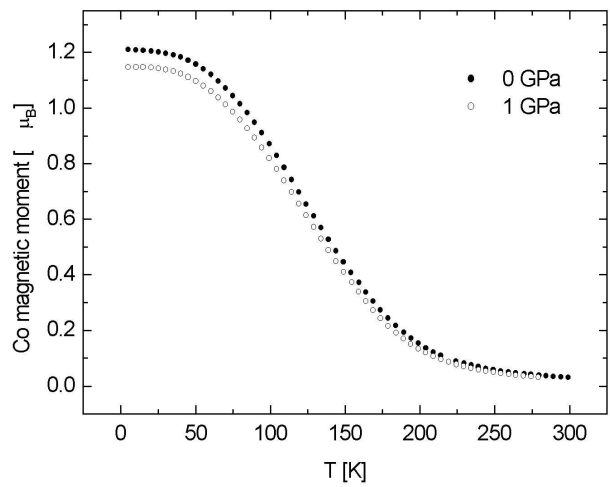

Fig.4: Temperature dependence of the $Co$ magnetic moment of $La_{0.82}Sr_{0.18}CoO_3$ in an external magnetic field of 2 T at ambient pressure and 1.0 GPa as deduced from magnetization measurements.

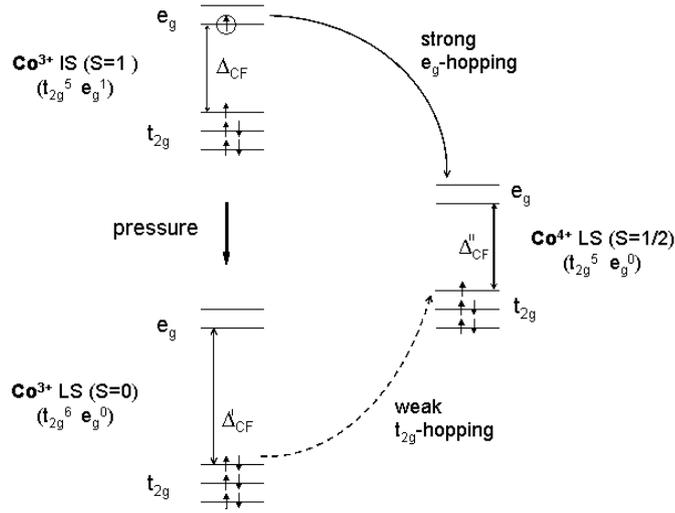

Fig. 5: Schematic representation of the interplay between spin-state transition and electron hopping in $La_{0.82}Sr_{0.18}CoO_3$: at ambient pressure conductivity occurs by hopping of an $e_g$ electron from IS $Co^{3+}$ to LS $Co^{4+}$. With increasing pressure, the crystal field splitting $\Delta_{CF}$ of IS $Co^{3+}$ increases $\left(\Delta'_{CF} > \Delta_{CF}\right)$ [25]. Consequently, the IS $Co^{3+}$ will be gradually depopulated and the ground state of $Co^{3+}$ is then predominantly a LS state. As a result, the strong $e_g$ hopping between $Co^{3+}$ and $Co^{4+}$ is strongly suppressed and only a weak $t_{2g}$ hopping remains.